\documentclass[preprint,12pt]{aastex}

\shorttitle{FTS Spectral Archive}
\shortauthors{Pilachowski et al.}

\begin{document}

\title{An Archive of Spectra from the Mayall Fourier Transform Spectrometer at Kitt Peak}


\author{C. A. Pilachowski}
\affil{Indiana University\\
Department of Astronomy SW319 \\
727 E 3rd Street \\
Bloomington, IN 47405 USA\\
cpilacho@indiana.edu}

\author{K. H. Hinkle}
\affil{National Optical Astronomy Observatory\\
950 N.Cherry Ave \\
Tucson, AZ, 85719 USA\\
khinkle@noao.edu}

\author{ M. D. Young\altaffilmark{3}, H. B. Dennis\altaffilmark{3}, 
A. Gopu\altaffilmark{4}, R. Henschel\altaffilmark{5}, and 
S. Hayashi\altaffilmark{6}}
\affil{Indiana University\\
University Information Technology Services\\
CIB\\
2709 E 10th Street\\
Bloomington IN 47401\\
youngmd@iu.edu, hbrokaw@iu.edu, agopu@iu.edu,
henschel@iu.edu, hayashis@iu.edu
}

\begin{abstract}

We describe the SpArc science gateway for spectral data obtained during the period from 1975 through 
1995 at the Kitt Peak National Observatory using the Fourier Transform Spectrometer (FTS)
in operation at the Mayall 4-m telescope. 
SpArc is hosted by Indiana University Bloomington and is available for public access.  
The archive includes nearly 10,000 individual spectra of more than 800 different 
astronomical sources including stars, nebulae, galaxies, and Solar System objects.   
We briefly describe the FTS instrument itself, and summarize the conversion 
of the original interferograms into spectral data and 
the process for recovering the data into FITS files.  
The architecture of the archive is discussed, and the process for retrieving data 
from the archive is introduced.  
Sample use cases showing typical FTS spectra are presented.

\end{abstract}


\section{Introduction} 

During the period from the mid-70s through the mid-90s, the Kitt
Peak National Observatory operated a Fourier Transform Spectrometer
(FTS) on the Mayall 4-m telescope.  The FTS produced spectra primarily
in the wavelength regime 1 to 5 $\mu$m (10,000 to 2000 cm$^{-1}$),
mostly at relatively high spectral resolution.  During this period,
before the development of infrared arrays, an FTS permitted efficient
observation of infrared spectra.  Nearly 10,000 sources were observed
with the 4-m FTS, many of them multiple times.  Targets included a
variety of normal and peculiar red giants, solar type stars, M
dwarfs, and variable stars, as well as nebulae and galaxies.  Early
type stars were occasionally observed to permit the removal of
telluric absorption lines.  Observations at high spectral resolution
were limited to a magnitude of K=4 or brighter, although some fainter
targets were observed at lower resolution.

Fourier transform spectroscopy has a number of advantages over
dispersive (grating) spectroscopy.  It is inherently a broad spectral
band technique.  The spectra in the 4 meter FTS archive generally
cover an atmospheric window in the J, H, K region and at least a
5\% (100 cm-1) bandpass in the thermal L and  M windows.  The
throughput is high and there is no scattered light.   The instrumental
function is well defined and the response is photometric.  The
wavelength scale is well defined without need for reference to a
dispersion relation and shifted by a small, constant amount from
absolute.  The spectral resolution is fully adjustable.  FTS instruments are
typically highly stable with time.  A dual input design is readily
implemented allowing first order removal of background, scintillation,
etc.  In summary the photometric accuracy and frequency calibration
of the spectra are limited by photon noise rather than instrumental
characteristics.  For these reasons this technique is very popular
in laboratory spectroscopy.  A wider view of the technique and
applications can be found in Davis et al. (2001).  However, the
non-dispersive nature of the technique is also a limitation.  With
the advent of infrared array detectors, Fourier transform spectroscopy
was at an increasing sensitivity disadvantage.  Multiplexing
disadvantage in spectrometers of high sensitivity is discussed by
Ridgway \& Hinkle (1988).

For bright objects we believe the spectra obtained with the 4-m FTS are unsurpassed.
The collection of 4-m FTS spectra forms a unique dataset that will not be duplicated with modern instrumentation.
More than 120 refereed papers using FTS data have been published since 1978, including nearly 
25 refereed papers since 2000\footnote{A list of refereed papers utilizing FTS data
is available at https://sparc.sca.iu.edu/index/papers.}.
The most highly cited works are seminal studies of CNO and their isotopes in red giants by 
Smith \& Lambert (1985, 1986, 1990) and Lambert et al. (1986), 
as well as studies of circumstellar material and mass 
flows in both young stellar objects and red giants with high mass loss rates (Hinkle et al. 1982).  
Saar and Linsky's 1985 paper on magnetic fields in the dM3.5e star AD Leonis 
conclusively demonstrated the presence of photospheric magnetic fields in dMe stars,
and Jorissen et al. (1992) used FTS spectra for the first measurements of the fluorine
abundance in stars other than the Sun.  
Mayall FTS data continue to be useful for a broad range of astrophysical questions, 
even 20 years after the FTS was formally retired from service.  
Recent papers include studies of dynamic model atmospheres and outflows, star 
forming regions, and stellar abundances.  Smith et al. (2013) used FTS spectra to provide 
a fundamental calibration of the APOGEE line list.  The excellent photometric quality of 
the spectra and high wavelength precision are ideal for atlases 
(Hinkle et al. 1995; Wallace \& Hinkle 1996; Wallace \& Hinkle 1997; Meyer et al. 1998; 
Wallace et al. 2000).

Here we describe the publicly available archive of spectra from the 4-m Mayall FTS. 
In Section 2 we summarize the characteristics of the instrument itself. 
Section 3 is devoted to a discussion of the data reductions and characteristics of the spectra. 
In section 4 we discuss the structure and features of the archive, and in Section 5 
we describe the contents of the archive and suggest use cases.
In Section 6 we summarize our work.

\section{The Mayall Fourier Transform Spectrometer} 

In the mid-1970s a prototype FTS with 10 cm travel and a design
based on a modified commercial instrument was installed at the f/160
coud\'{e} focus of the Mayall 4 m telescope on Kitt Peak.  The 4 m
telescope had been dedicated in 1973 and this was the first scientific
use of the coud\'{e} focus.  The prototype FTS was part of a collaboration
between D. Hall and S. Ridgway based on technology developed by J.
Brault for solar and laboratory spectroscopy.  The intent was to
provide a new capability for near-infrared spectroscopy of planetary
and galactic sources at the new telescope facility.  In 1978 September
the prototype FTS was replaced with a facility instrument (Hall et
al. 1979).  The facility Mayall FTS was significantly more versatile
and efficient than the prototype.  The maximum travel, which defines
the resolution, was 1.4 m.  The design permitted moderate to very
high resolution\footnote{Throughout this paper we will use wavenumbers
in cm$^{-1}$ units.  These are the natural units in Fourier
spectroscopy.  The spectra are transforms of intensity measurements.
The intensity is measured as a function of the path difference which
is typically measured in cm.  Hence the spectrum and resolution is
in inverse cm.} ($\sim$20 cm$^{-1}$ to 0.005 cm$^{-1}$) observations
from the red-optical through the mid-infrared $\sim$15000 cm$^{-1}$
to 700 cm$^{-1}$).  This facility FTS fully utilized in an astronomical
application the by then mature technology of Fourier spectroscopy.
A review of the FTS technique with emphasis on the application at
Kitt Peak\footnote{During the lifetime of the Mayall FTS, another
FTS operated at the McMath-Pierce solar telescope. The McMath FTS
was optimized for precision measurements of laboratory and solar
sources.} can be found in Brault (1985).  References to astronomical
applications of Fourier transform spectroscopy are given in Hall
et al. (1979).

\subsection{General Description}

Both of the 4 m FTS instruments were fed by the f/160 coud\'{e}
beam.  A key feature of the telescope, and in the use of the FTS,
was the rapidity with which the 4-m could be reconfigured between
f/8, f/16, and f/160.  This was done by moving the telescope far
to west and rotating sections of the top end to change the secondary
mirror, and took no more than 10 minutes.  Focal conversion was
especially advantageous to coud\'{e} use since time could be shared
between Cassegrain and coud\'{e} instruments.  The FTSs had a dual
beam design (star plus sky in one beam and sky in the other) that
allowed for sky cancellation.  This enabled near- and mid-infrared
spectroscopy during daylight hours in clear sky conditions.  The
FTS employed a rapid scan design with an interferogram recorded in
a few minutes.  Cancellation of sky was achieved by taking and
co-adding scans using an abba sequence.  As use of the facility FTS
matured, a large fraction of the observations took place during
daylight morning hours following nighttime use of an f/8 instrument.

Both the prototype and the facility FTSs were bench-mounted coud\'{e}
instruments directly fed by the coud\'{e} beam.  A photograph of
the facility FTS in the Mayall's east coud\'{e} room is shown in
Figure 1.  A schematic of the optical layout of the facility FTS,
copied with permission from the instrument description presented
by Hall et al. (1979), is shown in Figure 2.  The optical package
was held on a large optical table mounted on air supports.  The
individual optics were held by two axis adjustable Burleigh mounts.
The east coud\'{e} room is a closed room with a small window providing
a beam path from fifth mirror in the coud\'{e} train to the FTS.
The coud\'{e} number five mirror was servo-driven to track the beam
from the moving telescope onto the center of the FTS mirror 1 (Figure
2).  Mirror 1, an f/65 spherical mirror, reimaged the telescope
beam onto dual entrance apertures (at mirror 4 in Figure 2) and
placed the pupil of the entrance apertures at infinity.  The entrance
apertures were drilled into a spherical mirror and were either 3.5
arcseconds diameter for non-thermal or 2.7 arcseconds diameter for
thermal IR observations.  Note that at the time the FTS was used
the typical full width of the coude images was 3 arcseconds.  The
apertures were typically separated by 50 arc seconds.  These beams,
source plus sky in one beam and sky in the other beam, passed through
the FTS in opposite directions (APT1 and APT2 in Figure 2), which
resulted in first order sky cancellation.  Light reflected from
mirror 4 was directed to a TV for guiding.

The light that passed through the apertures was collimated by mirror
5 on the A side and an identical mirror on the B side and recombined
at the top half of a beamsplitter (6).  In Figure 2 the arrows show
the number of recombined beams.  The beamsplitter was a thick plate
of a optical material selected for the wavelength of the observation
with 50/50 reflection/transmission coatings on opposite sides of
the top and bottom and anti-reflection coatings opposite to the
coatings (Figure 3).  From the beamsplitter the light traveled to
two identical cat's eye assemblies (7-8-9-10-11 in Figure 2).  The
cat's eyes utilized a quartz front plate with a convex reflection
spot (9) between entrance apertures (7, 11) and quartz concave
primaries (8, 10).  At zero path difference the centers of curvature
of both the concave and convex surfaces were on the beamsplitter.
The result is that the input aperture was imaged on the cat's eye
secondary and the pupil re-imaged on the lower half of the beamsplitter.
This innovative cat's eye design is discussed further by Hall et
al. (1979).  The cat's eyes did not move symmetrically about zero
path.  The B side cat's eye was on rails and accounted for the major
travel.  The cat's eye on the A side was on a hinge mount and moved
rapidly to account for path errors.  The beams recombined at 12.
From there the beam went to camera mirrors (13 and the corresponding
mirror on the B side).  The final reflections were flats that sent
the light into dewars (14 on the A side and another on the B side)
containing single element detectors.  The overall number of reflections
involved was five in the telescope and eleven plus two beamsplitter
passes in the interferometer.  The FTS reflective surfaces were
silver with a ThF$_4$ overcoating.  The throughput at K was about
20 percent (including the atmosphere).  The modulation efficiency
at K was typically near 80 percent.

\subsection{Features Impacting the Spectra}

\subsubsection{Beamsplitters and Detectors }

The facility FTS optical design allowed operation from 0.3 to 30
$\mu$m by using interchangeable beamsplitters and detectors.  This
was facilitated by making the beamsplitter a plain-parallel plate
(Figure 3).  The design permitted beamsplitter changes with only
minor re-alignment of the FTS.  Beamsplitter changes could and were
carried out during an observing session.  The beamsplitters were
kinematically mounted and the realignment procedure (Simmons 1978)
could be carried out in less than 30 minutes.  By contrast the
McMath FTS beamsplitters were wedged and changed infrequently on a
scheduled basis.  The cost of the plane-parallel beamsplitter design
is that multiple internal reflections result in channel spectra.
The beamsplitter thickness and antireflection coatings reduced the channel
spectra.  However, with certain configurations the channel spectra
are prominent enough to be seen in the interferograms.  The fringing
was sometimes mitigated by removing the signal at the channel
frequency from the interferogram during data reduction.  Zeroing
the interferogram introduces some ringing.  Fringes from beamsplitter
channel spectra have a spacing of several per wavenumber.

For observations from 1.0 to 5 $\mu$m, two CaF$_{2}$ beamsplitters
were used depending on wavelength.  The detectors were InSb single
channel detectors operated in photovoltaic mode.  For longer
wavelength observations a NaCl beamsplitter was available and at
shorter wavelengths a fused silica beamsplitter was used.  These
were largely used for program with visitor supplied detectors.

Three sets of InSb detectors were available and optimized for
different signal levels.  Dewars A and B had LN2 cooled InSb detectors
with 1 mm diameters that were intended for broad band, bright sources
or thermal IR observations.  Dewars C and D were LHe cooled InSb
detectors of the smallest diameter available and tuned for faint
objects.  A third set of dewars, E and F, used 0.5 diameter InSb
detectors.  These were typically used at LHe temperature.  While
the normal mode of operation used a pair of detectors, it was
possible to operate with a single detector but with the loss of
half of the signal.  The detectors were always used in combination
with a cold blocking filter.  Changing the blocking filter required
opening the dewar.  Some near-IR band passes required a cold long
wave filter and a warm short wave filter.  The most common example
is the H-K filter which was a cold thick green glass filter to block
thermal IR combined with a warm 1.6 micron long pass filter.  Some
filters had fringing.  This will appear with periods greater than
a wavenumber.

Dichroic mirrors that could be used in place of the bending flats
(labelled 14 on the optical diagram) allowed, for instance, K and
H band spectra to be observed simultaneously with the low noise
detectors rather than with a single broad band filter and noisier
detectors.  When the dichroic was used the interferograms from both
sets of dewars were combined.  These spectra should not be confused
with broadband spectra in the archive.  Other dichroic combinations
were possible, for instance M band or L band and K.

N$_{2}$O cells were also available that could be placed either
behind the input apertures or in front of the dewar windows.  These
cells added lines to the K band spectrum as a velocity reference.

The signals from matched detectors at the two outputs were both
summed and differenced.  The sum removed the interferogram to leave
the instantaneous stellar flux.  This was displayed to provide
guiding and transparency information.  The differenced signal yielded
an interferogram with negligible DC offset.  The difference signal,
electrically filtered to the appropriate bandpass (see scan control
and scan parameters below), was ratioed to the electrically filtered
sum signal to eliminate amplitude modulation of the interferogram
due to guiding errors, seeing, etc.

\subsubsection{Scan Control}

The 4 m FTSs were rapid scan devices that moved the carriage
continuously during an observation.  Data was recorded with the
carriage moving both away from (forward) and toward (reverse) the
beamsplitter.  During the course of an observation the forward and
reverse scans were combined separately.  Hence, each observation
consists of two spectra.  Occasionally an instrumental problem would
occur that destroyed one or the other of the averaged scans so the
archives do have some single spectra.

A Zeeman frequency stabilized 633 nm (15798.003 cm$^{-1}$ in vacuum)
helium-neon laser was used as the path difference reference.  The
laser was mounted behind FTS mirrors 5 and 13 with the beams
propagating though tabs coated on the beamsplitter for 633 nm (Figure
3).  It was found that the detectors could see scattered non-laser
radiation when observing faint objects so a  notch filter was
installed before the laser light entered the FTS.  The reference
laser chopped between Zeeman components at a rate F$_{z}$ set by
an internal oscillator.  For the 4 m FTS F$_{z}$=1.800000 MHz.  The
two Zeeman laser components had different polarizations.  Polaroid
filters on each side of the beamsplitter separated the components.
They were optically remixed and detected by a silicon diode.  The
phase shifts of the signal relative to the oscillator monitored the
path difference.  Continuous phase-lock techniques were used to
achieve path difference position and velocity control.  For 15798.003
cm$^{-1}$ a fringe is 0.633 $\mu$m = 6.33$\times$10$^{-5}$ cm.  The
path difference was held within an RMS error of 2 milli-fringe RMS
error during constant velocity translation, i.e. the path difference
was held to 1 nm and the position of the moving cat's eye to half
this, 0.5 nm.  The control system was common with the McMath
solar/laboratory FTS and a detailed description is given by McCurnin
(1981).

\subsubsection{Scan Parameters -- M,K,S}

The carriage of the FTS scanned at a set constant rate.  The user
interface of the control system used three integer parameters to
control the carriage velocity, sample rate, and range of travel of
the carriage, M, K, and S.  These three parameters appear in the
headers and along with F$_{z}$ and $\sigma_{laser}$ were fundamental
to taking and reducing the data.  By definition the speed of the
moving carriage in laser fringes per second was F$_{z}$/M and the
sample rate in samples per second was F$_{z}$/K.

Three relations of interest can be derived from the definitions.  
If we define $f$ as the data frequency in hertz, then it follows that

\begin{equation}
{	F_{z} \over M }= {f \sigma_{laser} \over \sigma_{obs} }
\end{equation}

where $\sigma_{obs}$ is the wavenumber of the the observed data.
From the previous section, $\sigma_{laser}$=15798.003 cm$^{-1}$ and
F$_z$=1.800000 MHz, so

\begin{equation}
M = {114 \sigma_{obs} \over f } 
\end{equation}

The analog interferogram resulting from scanning the moving catseye
was electrically filtered to remove out of band noise.  Typically
one of two filters was used.  For bright or thermal IR objects the
smallest wavenumber part of the spectrum was placed at 400 Hz.  For
fainter objects the frequency response of the detectors was better
matched by scanning five times slower with the low cut off at 80
Hz.  Following these rules M was calculated for the wavelength
region to be observed.  Values associated with some common settings,
for instance the K band are 1140 for 400 Hz.  The analogue electrical
filters and the other control electronics were rack mounted in the
observing room next to the FTS room (Figure 4)

M/K is the number of samples per laser fringe.  This must be larger
than 0.5 to prevent under sampling the interferogram.  Thus the
high frequency edge of the electrical filters were 1000 Hz and 200
Hz.  The largest wavenumber that can be observed unaliased is given
by $\sigma_{laser}$M/2K. This of course had to be larger than the
blue edge of the optical filter.   Common settings for K and M were
900 and 1140.  K=900 produced a 2 KHz sample rate.  M = 1140 placed
the alias at 10005 cm-1.

There are a few observations that cover the entire J-H-K region.
A more common configuration was H-K or a single atmospheric window.
Since an FTS is not dispersive the noise from the single element
detector is distributed across the entire band pass Ridgway \&
Hinkle (1988).  In the thermal IR the background noise requires
narrow filters.  At M the filters were typically 100 cm$^{-1}$
broad.

A third scanning parameter sets the length of the scan.  The total
number of samples recorded was MS.  The total distance traveled in
units of reference laser fringes was KS.  The time to obtain a scan
is the distance traveled divided by the velocity, i.e.  KSM/F$_z$.

\subsubsection{Position of the Central Fringe}

The interferogram is a map of the interference with path difference.
This must be measured from zero path difference which is the middle
of the 'central fringe.'  The central fringe contains the lowest
frequency information regarding the overall shape of the spectral
bandpass, continuum, and zero point.  In practice the central fringe
was found in the data during the reduction.  The central fringe was
used to interpolate the interferogram so the output of the Fourier
transform was in one plane of conjugate space.  This required data
on either side of the central fringe.  Interferograms can be taken
with the central fringe in the middle or at one end.  The 'in the
middle' variety is called two sided and had the advantage that all
the data were converted into the spectral domain.  The central
fringe near one end was called one sided.  Typically approximately
ten percent of extra interferogram was scanned before the central
fringe.  While the ten percent was lost time, the advantage was
that the total scan time for the interferogram was cut approximately
in half.  This allowed more scans to cancel background.  The sidedness
of the interferogram is recorded in the headers.  The total path
difference gives the resolution so knowledge of the sidedness and
K and S is required to compute the resolution.

\subsubsection{Resolution versus Point Spacing}

A path difference of 1 cm results in a theoretical resolution of
1.0 cm$^{-1}$.  To obtain a path difference of 1~cm, the carriage
must move 0.5 cm since the light travels out and back from the
beamsplitter.  So 1 cm$^{-1}$ of theoretical resolution is a travel
of 0.5 cm/6.33$\times$10$^{-5}$ cm = 7899 fringes.  The path
difference is just one of several criteria that can be used to
define spectral resolution (Davis et al. 2001).  A definition of
resolution of more practical use is the path difference that produces
an instrumental profile with the resolution as the full width at
half maximum.  This is 9532 fringes for 1 cm$^{-1}$.  Since the FTS
instrumental profile results from the transform of a rectangular
function, i.e. the interferogram, the natural instrumental profile
is a sinc function.  The ringing side lobes of a sinc function can
be damped by apodizing, i.e. convolving the spectrum with a damping
profile, or alternatively multiplying the interferogram with a
weighting function.  A useful discussion and suggested apodizing
functions can be found in Norton \& Beer (1976).  Function 2 of
Norton \& Beer requires a path difference of 13345 fringes to produce
a FWHM of 1 cm$^{-1}$.  All the 4 m FTS archive data are unapodized.
For stellar spectra the sinc function profile will not be obvious.
However, we recommend apodizing the data from the facility FTS.  For 
most applications a Gaussian is acceptable.
Convolution with an anodizing function of appropriate FWHM should 
also be used to lower the spectral resolution if this is desired 
(see for example Wallace \& Hinkle 1997).

Interferograms were converted to spectra using the Fast Fourier
Transform (FFT) algorithm.  This requires a data set of length a
power of two.  In the reduction process zeros were added to the end
of the interferogram before it was transformed.  The effect of
adding zeros is to interpolate the spectrum.  \textit{As a result,
the point spacing of the spectra is not simply related to the
resolution. } The resolution must be computed by dividing the KS
product into the number of fringes for 1 cm$^{-1}$ for the line
profile you are using.  For instance for unapodized data and using
the FWHM definition, the resolution is 2*9532/KS for two sided
interferograms. The resolution will always be slightly lower than
the point spacing.

From the M, K, and S values it is possible to compute the point
spacing as well as the resolution.  The recipe is first to compute
the number of points, MS, but remember to divide by two if the
interferogram is two sided.  Then find the next power of two larger
than this number.  Take the power of two number of points and divide
it into the wavenumber of the alias, $\sigma_{laser}$M/2K,
by the power of two number of points.  This can be useful in
understanding the oldest archival data where header information is
missing.

The prototype FTS had an instrumental FWHM $\sim$10\% greater than
the theoretical value.  This resulted from instrumental apodization
associated with the vignetting of the pupil on the beamsplitter
with path difference.  The facility FTS yielded theoretical sinc
instrumental profiles.  The instrumental profile as indicated by
sharp terrestrial lines remained stable over the life of both the
prototype and facility instruments.  The cat's eyes were never
disassembled.  While the collimators were recoated the instrument
never underwent a major disassembly during its lifetime.

\subsubsection{Wavelength Scale Complications}

As discussed above the frequencies of all points in a spectrum are
expressed as multiples of the reference laser frequency.  The 4 m
FTS was not in vacuum enclosure so a compensation for the dispersion
of air was applied.  Furthermore, the reference laser and source
beam did not travel the same path through the FTS.  As a result the
step size for the reference beam is slightly different from that
of the source beam.  The shift in the wavenumber scale is small,
typically a fraction of a km s$^{-1}$.  This shift can be determined
by measuring the velocities of the telluric lines.  In some spectra
reference N$_{2}$O lines are also present.
Precision laboratory values for frequencies of lines in telluric and
N$_{2}$O spectra can be found in the HITRAN database
(Rothman et al. 2013; https://www.cfa.harvard.edu/hitran/).

\subsection{The Demise of the FTS} 

The magnitude limit for the 4 m FTS in the K band was about K=4 for
resolving power R = $\sim$20,000 with S/N~100 in 4 hours of observing.
The 4 m FTS was, in principal, able to observe all the sources found
by the two micron IRC survey (Neugebauer \& Leighton 1969), one of
the system specifications.  The low resolution K band limit was
about K=9 at resolving power R=$\sim$1000.  Broad band (K+H)
observations at R = 70000 and S/N=100 took about 4 hours at K=+1.
The brightest stellar sources at K have magnitudes near --4 so the
FTS could observe an impressive number of sources.  While far more
efficient than the previous generation of IR scanners, the FTS
nevertheless fell short of the sensitivity expected by optical
spectroscopists using photographic, let alone CCD, techniques.

The development of infrared array technology resulted in efficient
dispersive spectrographs for the infrared that are orders of magnitude
faster than the FTS (Ridgway \& Hinkle 1988).  Sensitivity is of
critical importance to modern astronomical instrumentation and the
FTS became obsolete.  As usage declined, there were other pressures.
The manufacturer of the A/D converter was no longer in business,
and replacement parts were not available.  Internal budget pressures
at NOAO resulted in no allocation of funds to keep the instrument
running.  Continuing changes in the 4 m telescope control system
were carried out without consideration of the effect on the aging
FTS software.  The final blow was the demand from engineering that
the coud\'{e} optics path be used for cooling air for the telescope
primary.  The Mayall FTS was turned off in June 1995.  Users of the
archive data from the last two years should be aware that stuck A/D
bits were common and there was no abba nodding.

\section{FTS Spectral Data}

\subsection{FTS Data Reductions}

Software (GRAMMY) to transform interferograms from the 4 m and solar
FTSs was written in the 1970s for the Kitt Peak CDC 6400 computer.
GRAMMY was a command driven package with a language of about 30
commands.  GRAMMY was ported to DEC computers after the CDC computer
was decommissioned but never rewritten.  Due to the limited memory
available at the time the software was written, revising the code,
while conceptually fairly simple, was a complex undertaking.  The
transform sizes were typically 2$^{18}$ to 2$^{20}$ data points.

Very limited real time reduction was available at the telescope.
Typically the only display available to the observer was of the
interferogram, although the S/N ratio could be approximated by using
the noise levels on the DC photometric signal.  Due to data storage
limitations only the two (one for forward scans and a second for
reverse scans) average interferograms were recorded for each
observation.

As noted above the forward and reverse scans were transformed
separately.  Comparing these pairs of spectra is a valuable guide
to the data quality.  In principal the data should differ only by
white noise.  Some common problems can be seen in the difference,
however.  Deviations from zero may indicate clouds during the
observation or distortion in the central fringe due to saturation.
60 cycle noise appears as evenly spaced bursts of noise.  The
relation of the noise bursts to their separation in wavenumbers can
be computed using the M and K parameters.  Guelachvili (1981)
provides a detailed discussion of distortions in FTS spectra.

After the interferograms were transformed, the spectra were trimmed
to remove regions outside of the filter band pass.  These wavelength
limits were input by the person reducing the data, are not constant
for the same blocking filter, and as a result sometimes spectral
information was lost.  The software also transformed 4096 points
around the central fringe to produce a low resolution spectrum.
The low resolution transform was archived with the full resolution
transform.  The limited computer resources available at the time
resulted in all the transforms being done on the main frame computer
in Tucson.  Thus it was possible to collect all the spectra into
an archive.  The archive was copied and combined multiple times.
Fortunately from the beginning multiple copies of the archive were
made and in spite of the poor quality of the recording media it was
ultimately possible to recover all the spectra ever transformed.
The full set of data occupied 63 9-track magnetic tapes.

Due to memory limitations the spectra were broken into 4096 point
blocks each containing the full header information.  Since the total
length of the transformed spectrum is always a power of two, breaking
the data into 4096 point = 2$^{12}$ point blocks was convenient.
Complex data were saved occasionally when the interpolation ('phasing')
of the interferogram was uncertain.  Changing the header information
was difficult and required editing the spectrum with REDUCER=DECOMP,
a program written by J. Brault for spectral analysis; header changes
were rarely made.

With the advent of personal computers, the 9-track tapes were stored
in memory.  The FTS data storage format was a pre-FITS format
invented by J. Brault.  This format was readable on CDC and SUN
computers but not easily decoded on other machines.  To read the
archived data on desktop computers, the archive was converted to
ASCII using DECOMP on Sun computers before those, too, were retired.
While standard DECOMP routines that had been in use for years were
used to convert the archive to ASCII a small digitization error is
noticeable if the wavenumber spacing is compared to the wavenumber
of the start of each block.

\subsection{Conversion to FITS Files} 

The 63 ascii ''card image'' files holding the full collection of
FTS spectra are formatted into sequential records of information
blocks and data blocks.  Each of the files contains multiple
observations.  Generally, each observation includes a low resolution
thumbnail and forward and backward scans at high spectral resolution;
for some spectra, the complex output of the transform was also
retained.  Each forward and backward scan in the ascii file is
broken into sub-scans of 4096 data points.  Each observation begins
with a big information block (BIB) with metadata associated with
the observation, and each sub-scan within the observation begins
with a small information block (SIB) that includes wavelength
information and scaling information pertinent to that sub-scan.

The process to convert data in an ascii file to FITS format began
with reading each record in the file and storing the metadata and
data values.  Since the observations were accumulated over twenty
years, the output of the original processing was not always consistent.
Each observation was checked by the software for inconsistencies
and examined in detail if inconsistencies were found.  As the file
was read, subsequent scans that were determined to be the forward
and reverse scans of the same FTS observation were matched.  Subsequent
records were determined to be associated if they have the same
target, scan number, and data type.  Subsequent records were
determined to be forward/reverse pairs if they have the same target
and have consecutive scan numbers.  The first scan in a pair is
identified as the forward scan.  Generally, when only one scan is
present, it is identified as a forward scan or backward scan depending
on whether its scan number is odd or even.  In some cases, the
metadata were ambiguous as to whether records were from the same
scan, or were a forward/reverse pair of the same scan.  In such
cases, the records were placed in different scans.  In some early
observations, scan numbers were not available in either the big or
small information blocks.  In these cases, the first scan was
assigned as the forward scan and the second scan as the backward
scan.  A history string in each FITS header specifies from which
input file and which records the data in that file come, as well
as any inconsistencies identified.

The spectral data, stored as ascii integers, were converted to
floating point using the scale factor and offset packing values
associated with each sub-scan.  The sub-scans associated with a scan
were concatenated to produce a single spectrum for each image
extension in the resulting FITS file, with the scan number as the
FITS extension.  In most observations, the forward scan was stored
in FITS extension 1 and the backward scan as FITS extension 2.

\subsection{Metadata}

Most observations are accompanied by complete metadata.
The original metadata available with FTS observations included observer-entered parameters
(e.g. the astronomical target) as well as information from the instrument and telescope computers.
We have, to the extent possible, preserved all metadata associated with each observation.
For some observations, communication between the telescope and instrument was not
operating normally, so for those observations, no telescope information is available.
The KPNO or NOAO program number is also included, when available.

Telescope metadata included current epoch coordinates, UT and integer Julian dates, 
and the beginning and ending time of observation (in seconds) for most observations obtained starting in 
November, 1976 (prior to that date, telescope communication was not available) and ending in 1993. 
Starting in January, 1994, through 1995, telescope coordinates
are not available for many of the observations.  
The starting and stopping sidereal time (in seconds), hour angle (in seconds), 
zenith distance (in radians), and air mass were also recorded.
These data were converted to standard notation and recorded using appropriate FITS keywords.

Telescope coordinates were precessed to J2000 and matched with SIMBAD sources to confirm
target identification and to obtain accurate J2000 coordinates to add to the FITS headers.  
This target confirmation is not possible for cases where telescope coordinates were not recorded.
In nearly all cases the observer-entered target name was confirmed.  
In a few cases, the telescope coordinates did not match the observer-entered target name.  
These spectra are identified in Table 1.  Users should examine these spectra
carefully to assure they are the correct object.
The exact pointing of many of the non-stellar objects is also uncertain, as are the specific 
identifications entered by observers, especially where the observer's identification is cryptic 
and telescope coordinates are unavailable.

Instrument parameters included the number of co-added scans; detector,
beamsplitter, and filter identifications;  scan direction; detector
gains, the point number of the central fringe, the laser and Zeeman
frequencies; the high and low frequency cuts for the interferograms,
the point number of the central fringe, and the FTS M, K, and S
parameters.

Data reduction parameters include the final dispersion parameters, the spectral resolution,
and the data units (all spectra are recorded and stored in vacuum wavenumbers).

Environmental metadata are also included when available, including
the atmospheric pressure, temperature, and water vapor pressure.
Wavenumbers are in vacuum and the environmental conditions
were used at the time of the data reduction to correct the wavenumber
scale.  We do not know how the wavenumber scale was calibrated for
spectra without environmental conditions.  Where records exist 20 C and 600 mm 
were the temperature and pressure inputs.  If wavelength is desired
this is generally expressed in air at STP.  There are several
different conversions, the most frequently cited is Edl\'{e}n (1953)
but this may lack the accuracy required.  Allende Prieto (2011)
summarizes recent information on wavenumber to wavelength conversion. 

In addition to the original metadata we have added, the J2000
coordinates noted above, photometric magnitudes in UBVRIJHK wavebands,
spectral types, and object classifications from SIMBAD to facilitate
searches by color, spectral type, and object class.

\section{SpArc Portal Design}

\subsection{SpArc Architecture}

The design of the SpArc web portal followed principles laid out by Gopu et al. (2014) 
and Young et al. (2015, 2016) for Scalable Compute Archive (SCA) systems developed 
through the Trident project at Indiana University. The archive employs 
a modern, intuitive web interface for searching, identifying, 
and viewing the spectra without the user directly accessing the data. 
PHP/Zend, Twitter/Bootstrap, AngularJS, and the HighCharts JavaScript 
library are used to produce the web interface elements, 
and these are connected by a REST API to backend data registration 
and retrieval services written in Python. 
A key element in the design is the reuse of components common 
to multiple projects and only modifying or developing those elements 
that are unique to meet the requirements of SpArc. 
Common components include metadata registration, data retrieval, 
authentication and authorization, and database querying.  
Components that were customized for SpArc include user interface elements,
metadata harvesting, and data visualization.

\subsection{SpArc Search Options}

Several search options enable the user to find and 
examine spectra of interest in the archive, as shown in Figure 5.
A name resolver will obtain J2000 coordinates from SIMBAD 
to carry out a search by position on the sky.  
The search tool defaults to a two arc second cone search, 
but larger diameter cone and box searches can be selected.  
The user also has the option to search using specific B1950 or J2000 equatorial, 
galactic, or ecliptic coordinates in either decimal or sexigesimal format.

Alternatively, the user can enter a specific object name, spectral classification, 
or object class to identify all targets matching the entered criteria 
(for example, all stars of spectral type K).  
Users who wish to identify all observations matching specific color or magnitude ranges, 
or observed within a specific date range, may also search on those criteria.  
The advanced search options allow the user to select spectra matching either any or 
all of the criteria (the default is to match all criteria).
Options to include low resolution scans and scans with uncertain dates can also be selected.
A batch search option allows users to find all observations matching a list of object names or
coordinates.
The final search results are displayed in a table.  
Table column headings can be customized to assist individual users to identify the spectra 
they need, and the table columns can be resorted as needed.

Once a list of spectra matching the desired criteria is found, users can review the individual spectra and associated metadata.  A table of all other spectra obtained on that date is also provided to assist
with identifying calibration spectra for telluric line division.
Users can download an individual spectrum or all of the spectra obtained on a given date, in FITS
format.
Forward and backward FTS scans are stored as FITS extensions.

\subsection{The SpArc Spectrum Viewer}

The quick-look spectrum viewer provided with the archive and shown in Figure 6
enables users to view both the full range 
of the spectrum and also to zoom in to display portions 
of the spectrum at full resolution.  
The spectrum viewer makes use of the HighCharts JavaScript 
library to display the spectral data, which was extracted 
from the individual FITS files during data registration 
and saved as a JSON-formatted file.  
The user can zoom into a narrow region of the spectrum by 
dragging the left and right limits to delineate the region of interest. 
The region can be shifted left or right using the slider provided. 
Display of either the forward or backward scan can be toggled, 
with both scans being overlaid by default.  
The wavenumber and intensity values appearing in the viewer 
are determined using a binned average of spectrum points 
to fit within the browser window. 
As the spectral region is narrowed, fewer and fewer points 
are included in each binned region until all data points are
displayed with the full resolution of the spectrum.

\section{Contents of the Archive} 


The archive includes nearly 900 individual (non-solar system) targets
including 780 individual stars, most with multiple observations.
The stars are distributed in spectral type as summarized in Table
2.  Observations of the Galactic center, variety of protostars and
Galactic star-forming regions, and several galaxies are also
available.  In addition to these targets, the dataset includes
multiple observations of Mercury, Venus, Mars, Jupiter, Saturn,
Uranus, Neptune, Comet IRAS, Titan, Io, and the Moon.

Because calibration observations are needed to remove telluric
lines, many observations of Vega, Sirius, and other hot stars were
obtained and are included in the archive.  These include 383 spectra
of Vega and 200 spectra of Sirius.  Mercury and the Moon were
observed in the thermal IR as telluric standards.  The telluric
features present in these infrared spectra provide information about
the Earth's atmospheric composition and physical condition during
the period 1975-1995.  These data are available for analysis to
monitor changes in the atmosphere over time, and to establish a
baseline against which future changes can be compared.  However,
users interested in research of this type may find the broader band
width, higher resolution, higher signal-to-noise data from the
McMath-Pierce FTS more applicable.

\subsection{Sample Use Cases}
 
The FTS Spectral Archive will support research on a variety of
science goals relating to the roles of mass loss, rotation, and
magnetic fields in the lives of stars, especially at the later
stages of stellar evolution.  The spectra will be of continuing
utility in calibrating and interpreting SDSS-III APOGEE spectra
collected by the Sloan Consortium (Smith 2013) and other infrared
spectroscopic surveys.  Furthermore, advances in stellar accretion
and mass models since the 1980s will lead to more rigorous analyses
of the many young stellar objects and proto-planetary nebula included
in the archive, helping to move those fields forward.  Studies of
stellar magnetic fields will also benefit from access to the FTS
spectra, since Zeeman splitting increases with wavelength, so that
fields are more detectable with infrared spectra than with optical
spectra.  In addition to the many features of CN, CH, CO, OH, NH,
HF, HCl, SiO and their isotopes, spectral features of interest in
the 1-5 $\mu$m regime include the Paschen, Brackett, and Pfund
series of hydrogen, many C I, Mg I, Si I lines and features of Al,
P, S, and K.  S-process elements are represented by Sr.  Finally,
the spectral coverage for some targets includes the He I 1.083
$\mu$m feature, useful for diagnosing chromospheric activity.

Three sample use cases are described below.

1.  A user wishes to identify all spectra of the RV Tauri star R
Sct in the K band.  SpArc returns a list of 20 observations of R
Sct obtained between November, 1982, and March 1983.  Two of the
spectra were observed at longer wavelengths (around 3000 cm$^{-1}$)
and one at a shorter wavelength (6000 cm$^{-1}$).  Eighteen
observations were observed in the 4000-4300 cm$^{-1}$ region.  Close
examination of the spectra identifies several that cover the specific
spectral region of interest.  After viewing and downloading each,
the user also identifies spectra of Vega and other calibration
observations taken on the same date, and downloads those as well.
Each observation includes a forward and backward scan, which the
observer averages to get a combined spectrum.  The user removes
telluric lines from each R Sct spectrum to produce the final spectra
needed for analysis.

2. A user wishes to model high spectral resolution, high signal-to-noise
line profiles of the He I 10830 \AA\ feature in early type stars
to constrain better the atomic parameters of the He I atom (Pryzbilla
2005).  The user identifies numerous spectra of early type stars in the
archive, sorts the spectra by wavenumber, and downloads all of the
spectra that include the 10830 \AA\ spectral region.  Multiple
observations available for some stars are examined to look for
variability in the feature, and combined to produce higher S/N ratio
spectra than possible with individual observations.

3. For late K and M giants where the peak thermal emission falls
in the infrared, temperatures are best determined in the H and K
bands.  The broad spectral coverage of the FTS in the H and K bands
gives access to lines in both the first and second overtone
vibration-rotation bands of the CO molecule, providing a sensitive
discrimination of temperature.  Temperatures derived using CO
features in late type giants can be compared directly with temperatures
determined from occultation and interferometric measurements or the
infrared flux method to fine-tune the temperature scale for such
stars, and establish the use of CO features to determine the
temperatures for fainter, more distant late-type giants observed
with modern infrared spectrographs.  A user can access the necessary
H and K band spectra using the archive.

\section{Summary}

We have described the Indiana University SpArc portal for spectra
obtained with the FTS on the Mayall 4-m telescope at Kitt Peak
during the years 1975-1995.  Both the instrument itself and data
reduction procedures have been documented, and we have described
the structure, contents, and use of the archive.  Three possible
use cases covering diverse science topics have been presented.  The
archive is freely available for scientific use, and can be accessed
at https://sparc.sca.iu.edu.  Users should acknowledge NOAO (see
our acknowledgments for an example) and the Indiana University SpArc
portal.

\acknowledgments

We are grateful to Catherine Gosmeyer for her contributions to the
original archive catalog, and acknowledge the important contributions
of J. Brault, D. Hall, and S. Ridgway in conceiving and building
the FTSs at Kitt Peak.  
We are also grateful to the referee for his/her prompt 
and enthusiastic review of our manuscript.
The data discussed in this article were
obtained with FTSs at the Mayall 4-meter Telescope at Kitt
Peak National Observatory (KPNO).  KPNO is a division of the National
Optical Astronomy Observatories, which is operated by the Association
of Universities for Research in Astronomy, Inc. under cooperative
agreement with the National Science Foundation.  This research has
made use of the NASA Astrophysics Data System Bibliographic Services,
as well as the SIMBAD database, operated at CDS, Strasbourg, France.
C.A.P. acknowledges the generosity of the Kirkwood Research Fund
at Indiana University.

\vspace{5mm}



\bibliographystyle{plainnat}

\begin{figure}
	\epsscale{.80}
	\plotone{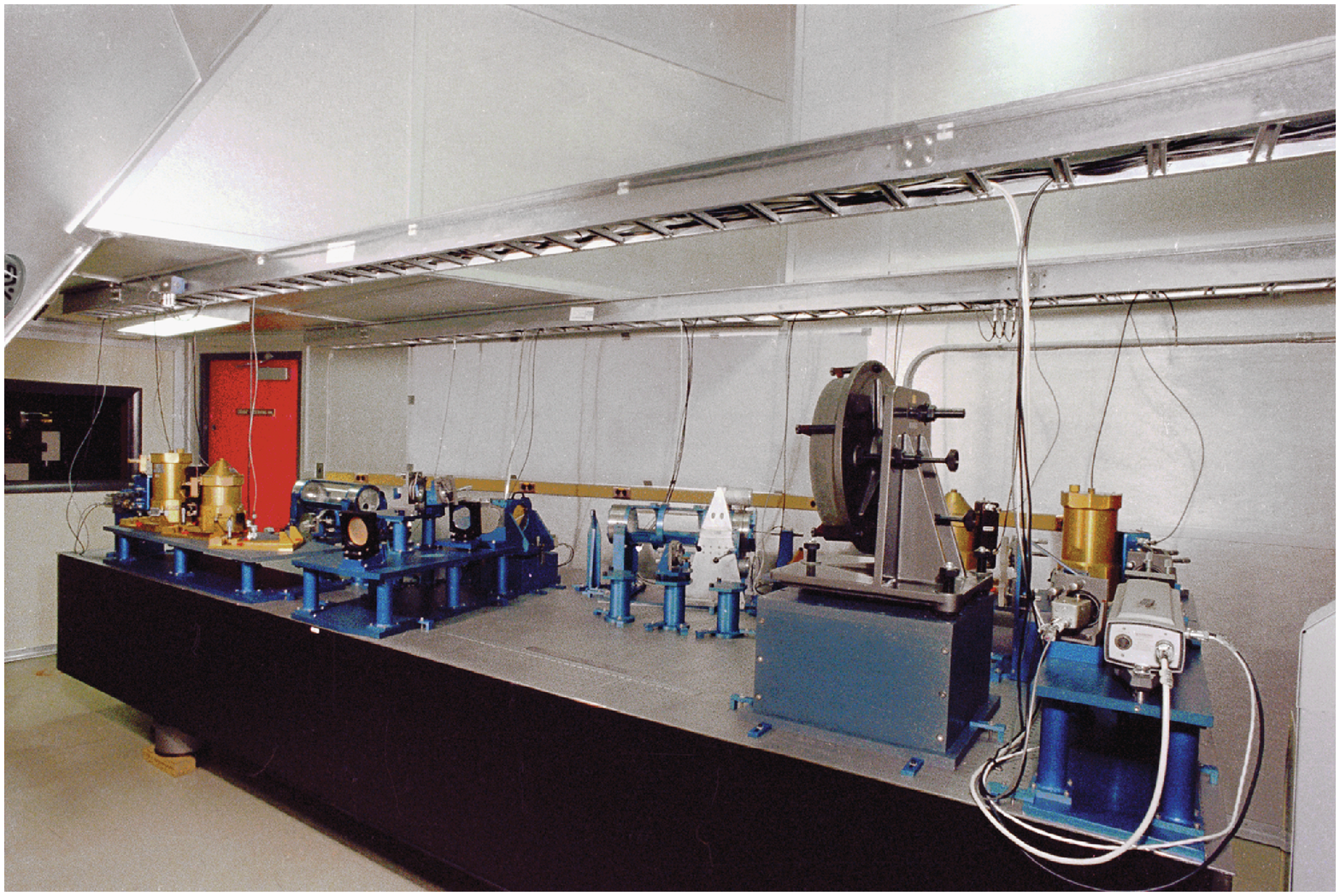}
	\caption{Photograph of the Mayall Fourier Transform Spectrometer in the east coud\'{e} room.  The number 5 coud\'{e}
	mirror was through the window at the left.  The number 5 coud\'{e} mirror tracked the f/160 beam onto the large mirror
	on the right.  The reference laser can be seen to the right of large mirror.  
		Image Credit: NOAO/AURA/NSF}
\end{figure}

\begin{figure}
	\epsscale{.80}
	\plotone{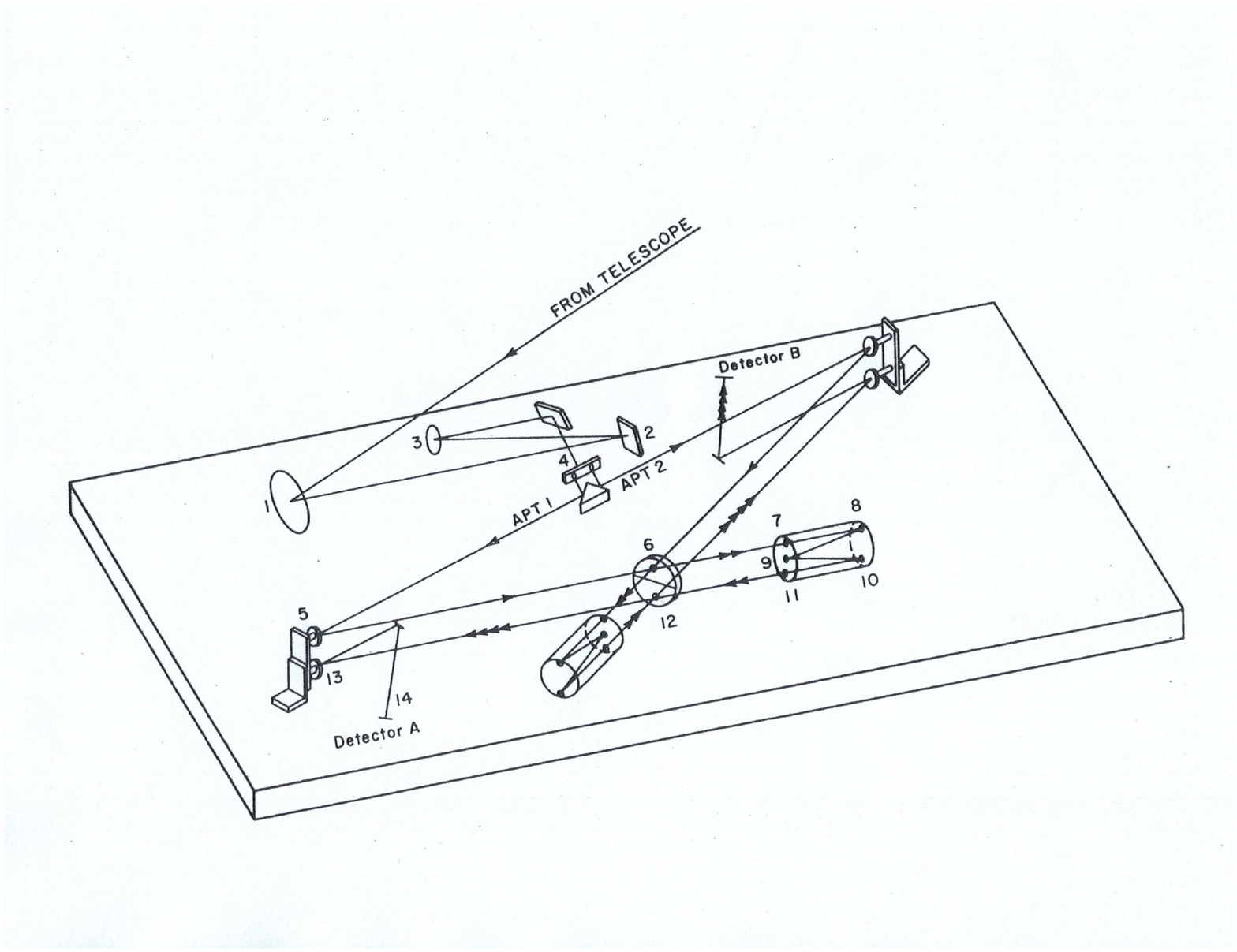}
	\caption{Optical layout of the facility FTS at the Mayall 4-m telescope at Kitt Peak National Observatory. Optical 
         components are discussed in Section 2.1.  This figure is Figure 2 from Hall et al. (1979, Proc. SPIE, 172, 121) 
         and is used with permission of D.N.B. Hall and the SPIE.}
\end{figure}

\begin{figure}
        \epsscale{.80}
        \plotone{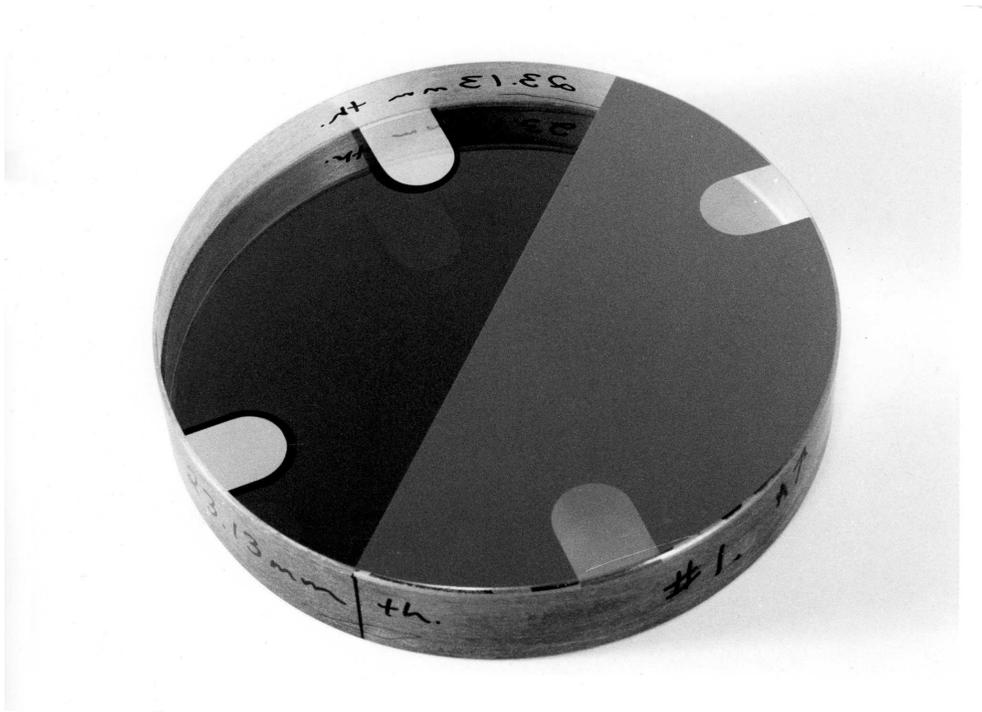}
        \caption{A beamsplitter from the facility FTS.  The substrate for this beamsplitter, designed for use at 4.6 $\mu$m, is a 
         23 mm thick CaF$_2$ plane-parallel plate.  Note the split coating design and the four tabs for the reference laser.  
         The opposite sides of the blank from the visible coatings were anti-reflection coated.
         Image Credit: NOAO/AURA/NSF}
\end{figure}

\begin{figure}
         \epsscale{0.80}
         \plotone{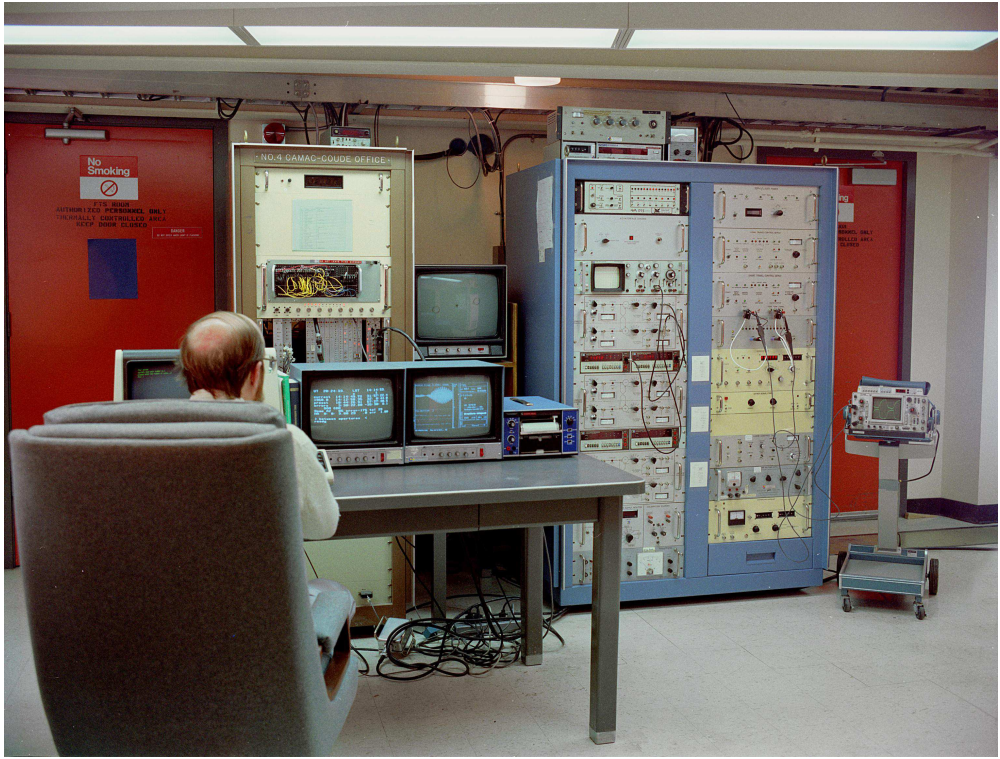}
         \caption{Photograph of the FTS observing room during an observing session.  KHH is the observer.  The right hand electronics 
         bay held the servo and laser 
         electronics.  The middle bay held the analog electrical filters for the interferograms and the A/D converter.  The left hand bay 
         was used for computer communication.  The central fringe of a K band interferogram appears on the monitor.
         Image Credit: NOAO/AURA/NSF}
\end{figure}


\begin{figure}
	\epsscale{.80}
	\plotone{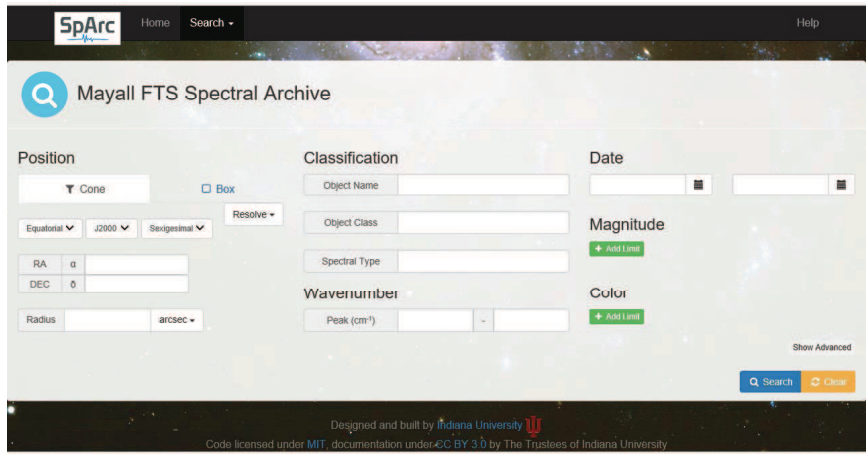}
	\caption{The SpArc search tool.}
\end{figure}

\begin{figure}
	\epsscale{.80}
	\plotone{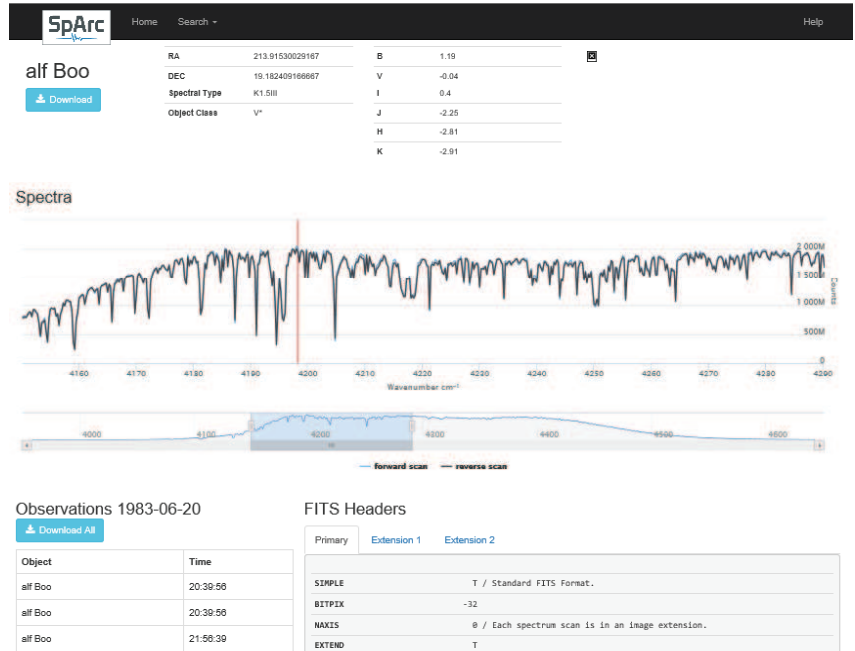}
	\caption{The SpArc spectrum display tool.}
\end{figure}

\begin{deluxetable}{ccc}
\tablecaption{Questionable Spectrum Identifications}
\tablewidth{0pt}
\tablehead{
\colhead{Date} & \colhead{Name Logged} & \colhead{Object at } \\
\colhead{of Obs.} & \colhead{by Observer} & \colhead{Tel. Coords}
}

\startdata 
10 May 1979 & MU U MA & alf Uma \\
24 Sept. 1980 & ZETA TAU 1 & eps Tau \\
23 Dec. 1980 & V VUL & T Vul \\
21 Feb. 1981 & ETA=62 CYG & Xi Cyg \\
13 Feb. 1984 & BETA TAU & eps Aur \\
10 Feb. 1985 & ALPHA BOO & \nodata \\
25 Jan. 1986 & MrK 321 & Mrk 231 \\
13 July 1992 & ST HER & OP Her
\enddata
\end{deluxetable}

\begin{deluxetable}{cccc}
\tablecaption{Spectral Types Represented in the Archive}
\tablewidth{0pt}
\tablehead{
\colhead{Spectral} & \colhead{Number of} & \colhead{Spectral} & \colhead{Number of}\\
\colhead{Type} & \colhead{Stars} & \colhead{Type} & \colhead{Stars}
}
\startdata
O  & 11 & K & 135 \\
WR & 7 & M & 257 \\
B & 65 & R & 7 \\
A & 65 & N & 8 \\
F & 47 & S & 30 \\
G & 88 & C & 60
\enddata
\end{deluxetable}

\end{document}